\documentclass[preprint,aps,prb,showpacs,superscriptaddress,amsmath]{revtex4}
\usepackage{graphicx}
\usepackage{dcolumn}

\usepackage{color}
\usepackage{ulem}

\begin{document}

\title{\textit{Ab initio} study of the CE magnetic phase in half-doped manganites: Purely magnetic versus double exchange description}

\author{Roland Bastardis}
\affiliation{Laboratoire de Physique Quantique, IRSAMC/UMR5626,
Universit\'e Paul Sabatier, 118 route de Narbonne, F-31062 Toulouse
Cedex 4, FRANCE} 
\affiliation{Department of Physical and Inorganic Chemistry, Universitat Rovira i Virgili, Marcel$\cdot$l\'{\i} Domingo, s/n. 43007 Tarragona, SPAIN}
\author{Coen de Graaf}
\affiliation{Department of Physical and Inorganic Chemistry, Universitat Rovira i Virgili, Marcel$\cdot$l\'{\i} Domingo, s/n. 43007 Tarragona, SPAIN}
\affiliation{Instituci\'o Catalana de Recerca i Estudis Avan\c{c}ats (ICREA) 
Passeig Lluis Companys 23, 08010, Barcelona, SPAIN}
\author{Nathalie Guih\'ery}
\affiliation{Laboratoire de Physique Quantique, IRSAMC/UMR5626,
Universit\'e Paul Sabatier, 118 route de Narbonne, F-31062 Toulouse
Cedex 4, FRANCE} 

\date{\today}

\begin{abstract}
The leading electronic interactions governing the local physics of the CE phase of half-doped manganites
are extracted from correlated \textit{ab initio} calculations performed 
on an embedded cluster.
The electronic structure of the low-energy states is dominated by 
double exchange configurations and O-2$p_{\sigma}$ to Mn-3$d$ charge transfer 
configurations.  
The model spectra of both a purely magnetic non-symmetric 
Heisenberg Hamiltonian involving a magnetic oxygen and two non-symmetric 
double exchange models are compared to the \textit{ab initio} one.
While a satisfactory agreement between the Heisenberg spectrum and the calculated 
one is obtained, the best description is provided by
a double exchange model involving 
excited non-Hund atomic states.  
This refined model not only perfectly reproduces the spectrum of the embedded cluster
in the crystal geometry, but also gives a full description 
of the local double-well potential energy curve of the ground state 
(resulting from the interaction of the charge localized electronic 
configurations) and the local potential energy curves of all excited states 
ruled by the double exchange mechanism. 
\end{abstract}
\pacs{75.47.Lx, 71.10.-w, 75.30.Mb, 75.30.Et}
\maketitle

\section{Introduction}

Although numerous theoretical and experimental studies have been devoted to 
the understanding of the various phase transitions occuring in doped manganites,\cite{jonker}
their electronic structure near half doping is still not fully understood.\cite{dagotto} 
La$_{0.5}$Ca$_{0.5}$MnO$_3$ is antiferromagnetic 
below the N\'eel temperature $T_N$ and paramagnetic for $T > T_{CO}$, where CO stands for charge-ordered. 
For intermediate temperatures $T_N <T< T_{CO}$ two different electronic stuctures have been proposed. 
In the first one, a charge order prevails and the magnetic phase exhibits a CE-type ordering of the magnetic 
moments.\cite{wollan,radaelli} In the second electronic structure,\cite{daoud} the holes are trapped in pairs of equivalent Mn ions of 
intermediate valence state Mn$^{3.5+}$ resulting from the double exchange (DE) interaction between the two limit 
Mn$^{3+}$O$^{2-}$Mn$^{4+}$ and Mn$^{4+}$O$^{2-}$Mn$^{3+}$ electronic configurations. The latter electronic structure 
is referred as a Zener polaron order in reference to Zener who first proposed the DE mechanism.\cite{Zen} 

\textit{Ab initio} calculations confirm the existence\cite{coen} of a charge-ordered phase (partial localization of the holes on one type of Mn ions) 
in the crystallographic structure of Radaelli \textit{et al.}\cite{radaelli} 
while a polaronic order\cite{coen,nous1,nous2,nous3} (delocalized holes between pairs of Mn ions) is obtained for 
the crystallographic structure of Daoud Aladine \textit{et al.}\cite{daoud}
It is worth to note that in both cases correlated \textit{ab initio} calculations point out that there is a rather strong O to Mn charge transfer, resulting in a partial localization of the holes on the bridging oxygens.  
UHF and DFT periodic calculations\cite{var2,var3,var3a} on La$_{0.5}$Ca$_{0.5}$MnO$_3$ even lead to a dominantly Mn$^{3+}$O$^-$Mn$^{3+}$ 
ground state charge distribution. 

The impact of this partial localization of the holes on the oxygen sites 
has recently been studied in the peculiar case of a Zener polaron i.e. using the 
crystal structure of Daoud Aladine \textit{et al.}\cite{daoud} The two Mn ions in the Zener polaron have an almost identical coordination, and hence, several symmetric models have been derived and their solutions contrasted with \textit{ab initio} results.\cite{nous1,nous2} 
It has been shown that i) a truncated-Hubbard in which all the dominant configurations 
are treated variationally can be mapped both on the Heisenberg and the usual 
DE model owing to the mixing of the bridging oxygen and Mn ions orbitals.\cite{nous2} 
ii) the energies of the eight lowest states of the Heisenberg and 
the usual DE models are analytically identical and in 
qualitative agreement with those of the \textit{ab initio} spectrum.\cite{nous2,nous3} 
iii) a significant improvement 
is observed when the non-Hund excited atomic states are explicitely introduced in the modelization.\cite{nous1} 
The remarkable accuracy of the resulting model spectrum finally shows that
the Zener polaron physics is actually ruled by 
a refined DE mechanism where non-Hund atomic states play a non negligible role. 

The present paper focuses on the local electronic structure of the charge-ordered crystallographic structure of La$_{0.5}$Ca$_{0.5}$MnO$_3$. 
For this purpose we have studied an embedded cluster containing two corner-sharing MnO$_6$ octahedra using the crystal structure 
determined by Radealli \textit{et al.}\cite{radaelli} The two Mn ions in the cluster have different coordination spheres. One site is Jahn-Teller distorted, while the other has six (almost) equivalent, rather short Mn-O distances. The two sites are commonly interpreted as occupied by Mn$^{3+}$(3$d^4$) and Mn$^{4+}$(3$d^3$), respectively.
According to the position of the holes, different electronic orders are expected: 
\begin{itemize}
\item If the holes are localized on the Mn sites, the resulting seven 
unpaired electron system is of mixed valence nature. The DE mechanism 
induces a resonance between Mn$^{3+}$O$^{2-}$Mn$^{4+}$ and 
Mn$^{4+}$O$^{2-}$Mn$^{3+}$ which in that peculiar case will result 
in a charge ordering of the electronic structure i.e. a partial localization 
of the e$_g$-like extra electron (or hole) on one Mn ion Mn$^{3.5+\delta}$O$^{2-}$Mn$^{3.5-\delta}$.
The t$_{2g}$-like electrons are unpaired and strongly localized on each ion.

\item If the holes are localized on the bridging oxygen, the corresponding charge distribution
suggests a dominant purely magnetic local order (Mn$^{3+}$O$^{-}$Mn$^{3+}$) in which the cluster 
would be a ferrimagnetic entity involving a magnetic oxygen and therefore nine unpaired electrons. 
The model Hamiltonian that provides a relevant description of such a local 
electronic order is a Heisenberg Hamiltonian involving non-symmetric exchange integrals between each Mn ion and the magnetic oxygen. 
\end{itemize}

\noindent In order to decide which Hamiltonian is most appropriate to describe the local electronic structure of this non-symmetric two-center cluster, we compare the different model
spectra with the \textit{ab initio} one. 
The here-used procedure consists in extracting the local 
effective interactions of the model Hamiltonians from 
calculations performed with the exact electronic Hamiltonian. 
For this purpose, the low energy spectrum of an embedded cluster 
of the material is studied by means of correlated \textit{ab initio} 
calculations. The subsequent extraction 
of the model parameters makes use of the effective Hamiltonian theory of Bloch.\cite{bloch,cloi} 
The comparison of the resulting model spectra to the \textit{ab initio} one 
provides rational arguments to determine the most appropriate model Hamiltonian.

Section II is devoted to the presentation of the embedded cluster 
computational procedure. The Heisenberg and DE models 
as well as their analytical solutions are derived in section III. 
Finally, the comparison of the model spectra with the
\textit{ab initio} one is discussed in the fourth section.

\section{\label{cluster_description}Description of the cluster and computational procedure}

\begin{figure}[b]
\centerline{\rotatebox{0}{\includegraphics[scale=0.5]{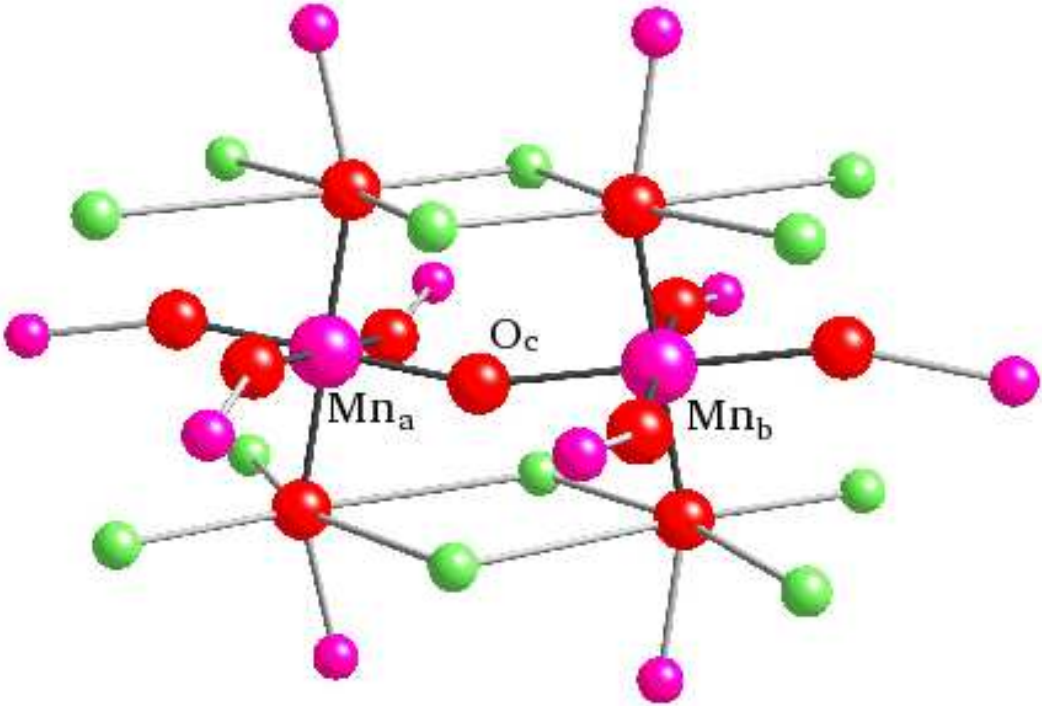}}}
\caption{\label{cluster} (color online) Mn$_2$O$_{11}$ cluster embedded in Mn and (La,Ca) total ion potentials (TIP's).}
\end{figure}

The geometry of the Mn$_2$O$_{11}$ cluster is derived from the 20K modulated structure of La$_{0.5}$Ca$_{0.5}$MnO$_3$ reported in Table I of Ref. \onlinecite{radaelli}.
Two different types of Mn sites can be distinguished in the material, the mean radius of their 
coordination spheres are respectively 1.971 \AA\ and 
1.916 \AA.
The distances between a bridging oxygen and the two distinct Mn sites are 
1.915 \AA\  for 
Mn$_\mathrm{a}$--O$_\mathrm{c}$ and 2.068 \AA\ for Mn$_\mathrm{b}$--O$_\mathrm{c}$. 
The cluster is represented in Fig. \ref{cluster} and is constituted of two different Mn ions 
in a distorted octahedral environment 
bridged by O$_\mathrm{c}$, and ten external oxygen atoms. 
Axis are chosen such that the origin of the coordinates is at half distance between the two Mn sites, 
$z$ is the intermetallic axis and the bridging O atom (O$_\mathrm{c}$) lies on the $xz$-plane.

To reproduce the crystal environment, the cluster is embedded in a set of optimized point charges 
which accurately reproduces the Madelung potential of the crystal. 
Total Ion Potentials (TIP) have been used to represent the Mn, La and Ca cations coordinated to the oxygen atoms of 
the cluster (see Fig. \ref{cluster}). Additional information can be found in Ref. \onlinecite{coen}, 
where it is shown that this embedding provides reliable results on the ground state 
of the system. 

The electronic structure of the cluster has been studied using the complete active space second-order perturbation theory (CASPT2) method\cite{caspt20,caspt21,caspt22,caspt2-rev} implemented in the \textsc{molcas} package.\cite{molcas} The orbitals are expanded with
extended basis sets of ANO type:\cite{ano,ano2} (5$s$, 4$p$, 3$d$, 1$f$) for Mn atoms, 
(4$s$, 3$p$, 1$d$) for the bridging O and (4$s$, 3$p$) for external O. 
The CASPT2 method, which has successfully been applied to the study of many 
spectroscopic problems, introduces both non-dynamical and dynamical correlation effects.
In a first step, a multireference wave function is optimized by the complete active space self-consistent field (CASSCF) procedure. This $N$-electron wave function includes the main 
non-dynamical electron correlation. 
The here-considered active space contains all electronic configurations 
obtained by distributing nine electrons in the ten $3d$ orbitals of the two Mn ions and 
the 2$p_{\sigma}$ orbital essentially localized on the bridging oxygen. 
Accurate relative energies of the ground state and lowest excited electronic states are computed by introducing the dynamical electron correlation in a subsequent second-order perturbation treatment applied to the zeroth-order CASSCF wave function.

Using the \textit{ab initio} spectrum and the effective Hamiltonian theory of Bloch,\cite{bloch,cloi} it is possible to extract 
the main electronic interactions of the different models.  
Since the number of electronic interactions is
smaller than the number of electronic states described by the model space, the value of the effective interactions is fitted to optimally reproduce the \textit{ab initio} spectrum. The mean error per state can be evaluated with the following expression:
\begin{eqnarray} \label{epsilon} \epsilon=\frac{100 \times \sum _i |E^{AI}_i-E^{M}_i|}{N \times \Delta E^{AI}}
\end{eqnarray}
where $N$ is the number of states described by the model Hamiltonian,
$E^{AI}$ and $E^{M}$ are respectively the \textit{ab initio} and model energies
of state $i$, and $\Delta E^{AI}$ the \textit{ab initio} spectrum width.

Both the comparison of the so-obtained model spectra to the \textit{ab initio} one and the size of the error per state give rational arguments 
to discriminate between different modelizations.

\section{\label{models}Description of the different models}

\subsection{The Heisenberg model}

If the holes are essentially localized on the oxygens O$_c$ then the wave functions of the lower energy states are dominated by 
electronic configurations with nine unpaired electrons. This is compatible with 
a Mn$_\mathrm{a}^{3+}$--O$_c^-$--Mn$_\mathrm{b}^{3+}$ electronic struture 
and the appropriate model Hamiltonian is the Heisenberg Hamiltonian. 
In a non-symmetric cluster, its specific form as a function of the exchange integrals between the three magnetic centers is:
\begin{equation}
\mathbf{H}^{Heis}= -J_1 \mathbf{S}_{Mn_\mathrm{a}}\cdot \mathbf{S}_{O_\mathrm{c}} -J_3 \mathbf{S}_{Mn_\mathrm{b}}\cdot \mathbf{S}_{O_\mathrm{c}} -J_2 \mathbf{S}_{Mn_\mathrm{a}}\cdot \mathbf{S}_{Mn_\mathrm{b}},
\end{equation}
where $J_1$ is the magnetic interaction between Mn$_a$ and O$_c$, 
 $J_3$ the magnetic interaction between Mn$_b$ and O$_c$, and 
$J_2$ the magnetic interaction between the two Mn centers.
The difference between the $J_1$ and $J_3$ exchange integral 
values reflects not only the geometrical asymmetry (the Mn$_a$--O$_\mathrm{c}$ distance is shorter than the Mn$_\mathrm{b}$--O$_\mathrm{c}$ distance) but also any possible asymmetry in the electronic structure.

The model space of this Hamiltonian is spanned by the products of the magnetic sites ground states. 
In the considered case, each Mn atom has four unpaired electrons coupled 
to a quintet ground state.
The bridging oxygen O$_\mathrm{c}$ has one unpaired electron in a local 
$2p_{\sigma}$-like orbital leading to a doublet ground state.
An analytical solution of the energies of the eigenstates of the 
Heisenberg model space has been derived for this simple case. 
Its expression is:
\begin{equation}
\label{magn_dets}
E^H(S,\pm) = \pm\sqrt{[S^H_{max}(S^H_{max}+1)-S(S+1)]x^2+\beta^2\left(S+\frac{1}{2}\right)^2} \nonumber \\
           + J_2[S(S+1)-S^{H}_{max}(S^{H}_{max}-1)]
\end{equation}
\noindent where $x=J_1-J_3$ and $\beta=J_1+J_3-2J_2$. $S^H_{max}$ is the maximal value 
that $S$ can take in the Heisenberg model. $S^H_{max}=9/2$ in the here-considered cluster.
The mean energy of the octet states is taken as the zero of energy.

\subsection{\label{model2}The usual non-symmetric double exchange model}

With the holes localized on the Mn ions, the $N$-electron wave functions of the lower states are dominated by spin configurations 
presenting seven unpaired electrons, 
i.e. a Mn$_a^{3.5+\delta}$--O$_c^{2-}$--Mn$_b^{3.5-\delta}$ 
(closed shell oxygen) electronic structure. 
The cluster is of mixed valence nature and the DE model 
should accurately reproduce its physics.

In a charge-ordered phase, i.e. with $\delta \neq 0$, the electronic structure of the cluster is such that the 
extra electron (or hole) is partially localized on one of the two Mn ions. 
The nuclear relaxation resulting from this partial localization 
results in an enlargement of the coordination sphere (an increase of the average Mn-O distance) of the Mn 
ion bearing the extra electron and a contraction of the coordination 
sphere of the Mn ion bearing the hole. 
Piepho, Krausz and Schatz \cite{PKS} proposed a vibronic coupling model
to describe this phenomenon in mixed-valence compounds.
The relevant dimensionless coordinate in this so-called PKS model is $q$, 
which describes the nuclear relaxation during the electron transfer process.  

In the PKS model q corresponds to the antisymmetric combination 
of the local stretching mode of the coordination spheres. The here-considered coordinate q
also includes the Jahn Teller distorsion, since the cluster is studied 
at the experimental geometry which is Jahn-Teller distorted.

Let us call $\Phi _a$ and $\Phi _b$ the electron localized diabatic states corresponding 
to functions in which the extra electron is localized on Mn$_a$ and Mn$_b$ respectively.
In the usual DE model these functions are products of the Mn ions 
ground state wave functions, i.e. in the $m_s$=7/2 subspace 
$\Phi_a=Q^i_a \cdot Q_b$ and $\Phi_b=Q_a \cdot Q^i_b$ where 
$Q^i$ is a quintet state and $Q$ stands for a quartet state.
Their energetic dependence to the coordinate $q$ is usually assumed to be quadratic, so that the potential energy curves of $\Phi_a$  and $\Phi_b$ 
obey the equations E$_a$= $\Delta (q^2+2q)$ and E$_b$=$\Delta (q^2-2q)$ respectively,  
where $\Delta $ is the curvature of the diabatic spin states potential
energy curves as functions of $q$. The PKS Hamiltonian expressed in the $m_s$=7/2 subspace is :
$$\left(\begin{array}{cc}
\Delta (q^2+2q) & t \\
t & \Delta (q^2-2q)
\end{array}\right)$$
with $t$ the hopping integral of the extra electron between the 
two metal centered orbitals ($e_g$-like in the present case). Its diagonalization generates the two adiabatic octet states $O_+=N(\Phi_a+\Phi_b)$ and $O_-=N(\Phi_a-\Phi_b)$ of energy
\begin{eqnarray}
E^{PKS}(7/2,\pm)=\Delta q^2\pm\sqrt{4(\Delta q)^2+t^2}
\end{eqnarray}
In a non-symmetric and therefore charge-ordered system 
the potential energy curve of the ground state $O_+$
as a function of the coordinate $q$ presents a double well. The corresponding minima are
associated to the two non-symmetric geometries and the Mn$^{3.5-\delta}$-O$^{2-}$-Mn$^{3.5+\delta}$ or  Mn$^{3.5+\delta}$-O$^{2-}$-Mn$^{3.5-\delta}$ charge-ordered electronic configurations. 

The coupling between the lower $m_s$ components of the local ground states $\Phi _a$ and $\Phi _b$ generates states of lower spin multiplicity, 
namely two sextet $S_-$ and $S_+$, two quartet $Q_-$ and $Q_+$, and 
two doublet $D_-$ and $D_+$ states. 
In the general case of a non-symmetric homonuclear bimetallic complex, the eigenenergies
of the model are analytically known\cite{gir1,gir2,papa} and are given by the expression:
\begin{equation}
\label{Zener1_equation}
E^{DE}(S,\pm)=\pm\sqrt{4(\Delta q)^2+B^2\left(S+\frac{1}{2}\right)^2}+J[S(S+1)-S^{DE}_{max}(S^{DE}_{max}+1)]
\end{equation}
where $S^{DE}_{max}$ is the maximal value that $S$ can 
take in the DE model, $S^{DE}_{max}=7/2 $ in the considered cluster. 
$B=\frac{t}{S^{DE}_{max}+\frac{1}{2}}$ is the leading interaction of the model. $t$ favors a ferromagnetic order, in other words, it accounts for the appearance of a high-spin ground state.
$J$ is the overall exchange integral which describes the electronic circulation 
in the other open shells, here the $t_{2g}$-like orbitals. 
This Heisenberg contribution is generally antiferromagnetic and
has been introduced by Girerd and Papaefthhymiou.\cite{gir1,papa} It causes 
a significant stabilization of the mean energies of the 
low and intermediate spin states with respect to the mean energy of 
the highest spin states.
The zero of energy is placed at the mean energy of the highest spin states.

The values of $t$, $J$ and the product $\Delta q$ are determined by minimizing $\epsilon$, defined in Eq. \ref{epsilon}.

The expression of the variable $q_{min}$ at the minimum of the ground state energy 
can be obtained from the zero of the first derivative of the energy:
\begin{eqnarray}
q_{min}&=&\frac{2\Delta q}{\sqrt{4(\Delta q)^2+t^2}}
\label{extracq}
\end{eqnarray}
Knowing the product $\Delta q $ and the value of $q$ at the minimum,
it is possible to determine the value of the curvature $\Delta$ and to draw
the full potential energy curves as a function of $q$. 

The hopping parameter can also be extracted directly from the energies and wave functions of the two octet states without fitting the \textit{ab initio} spectrum to a model Hamiltonian. To validate our numerical fitting procedure, we compare the $t$-values obtained in both procedures.
The symmetrically orthonormalized projections of the adiabatic octet state wave functions onto the model space are:
\begin{eqnarray}
|\Psi_1\rangle&=&cos \phi|\Phi_a\rangle + sin \phi|\Phi_b\rangle,\nonumber \\
|\Psi_2\rangle&=&-sin \phi|\Phi_a\rangle+cos \phi|\Phi_b\rangle.
\end{eqnarray}
where $\frac{\pi }{4}-\phi$ characterizes the charge localization.
From these expressions, the des Cloizeaux\cite{cloi} formalism allows 
us to extract the hopping integral as follows:
\begin{eqnarray}
\label{extractt}
t&=&sin \phi cos \phi [E(O,-) -E(O,+)]
\end{eqnarray}

\subsection{Explicit treatment of the non-Hund excited atomic spin states}

The explicit consideration of the excited atomic non-Hund states in the 
DE model was introduced by Anderson and Hasegawa\cite{And} in the description of a
symmetric (i.e. fully delocalized) cluster.
In comparison to the usual DE model, the Anderson-Hasegawa  
model does not only consider the products of local ground states. 
Its model space is extended to products of an atomic ground state on one Mn 
ion and a single excited non-Hund atomic state on the other Mn ion. 
The so-obtained functions interact with the model space of the 
usual DE model through a term proportional to the hopping 
integral $t$. 
The main contribution of a variational treatment of these functions
is antiferromagnetic. This means that the energies of the highest spin states 
are not affected, while the intermediate and low spin states are stabilized by 
the non-Hund state contribution. 
As already shown in symmetric dimers,\cite{nous1,nous2,nous3,nini,rere,david}
the best results are obtained using this refined DE model in combination with the antiferromagnetic contribution of Girerd-Papaefthymiou through a 
magnetic exchange integral $J$. 
Assuming that the potential energy curves of the electronic states formed by a product 
of a non-Hund state and an atomic ground state are parallel 
to those of the products of the atomic ground states, 
it is possible to generalize this refined model to the study of 
non-symmetric clusters. 
Calling $\delta_H$ the relative energy of the non-Hund state,
one obtains the following energy expression of the DE states:
\begin{eqnarray}
E^{DE}_{NH}(S,\pm)&=&\frac{1}{2}\left[\delta_H-\sqrt{\delta_H^2+16(\Delta q)^2+4t^2\pm4\delta_H\sqrt{4(\Delta q)^2+B^2\left(S+\frac{1}{2}\right)^2}}\;\right]\nonumber \\
             &+&J[S(S+1)-S^{DE}_m(S^{DE}_m+1)]
\end{eqnarray}
Again, the electronic interactions $t$, $\delta_H$, $J$ and the product 
$\Delta q$ are numerically optimized such that the model spectrum optimally fits  
the \textit{ab initio} one. 
Since the contribution of the non-Hund states is zero in the highest spin states 
and therefore in the ground state, the extraction of $q_{min}$ (and then $\Delta$) is performed using 
Eq. \ref{extracq}.

\section{\label{results}Contrasting the models to \textit{ab initio} results} 

\subsection{\label{results_a}Comparison of the spectra}

\begin{figure}
\centerline{\rotatebox{-90}{\includegraphics[scale=0.6]{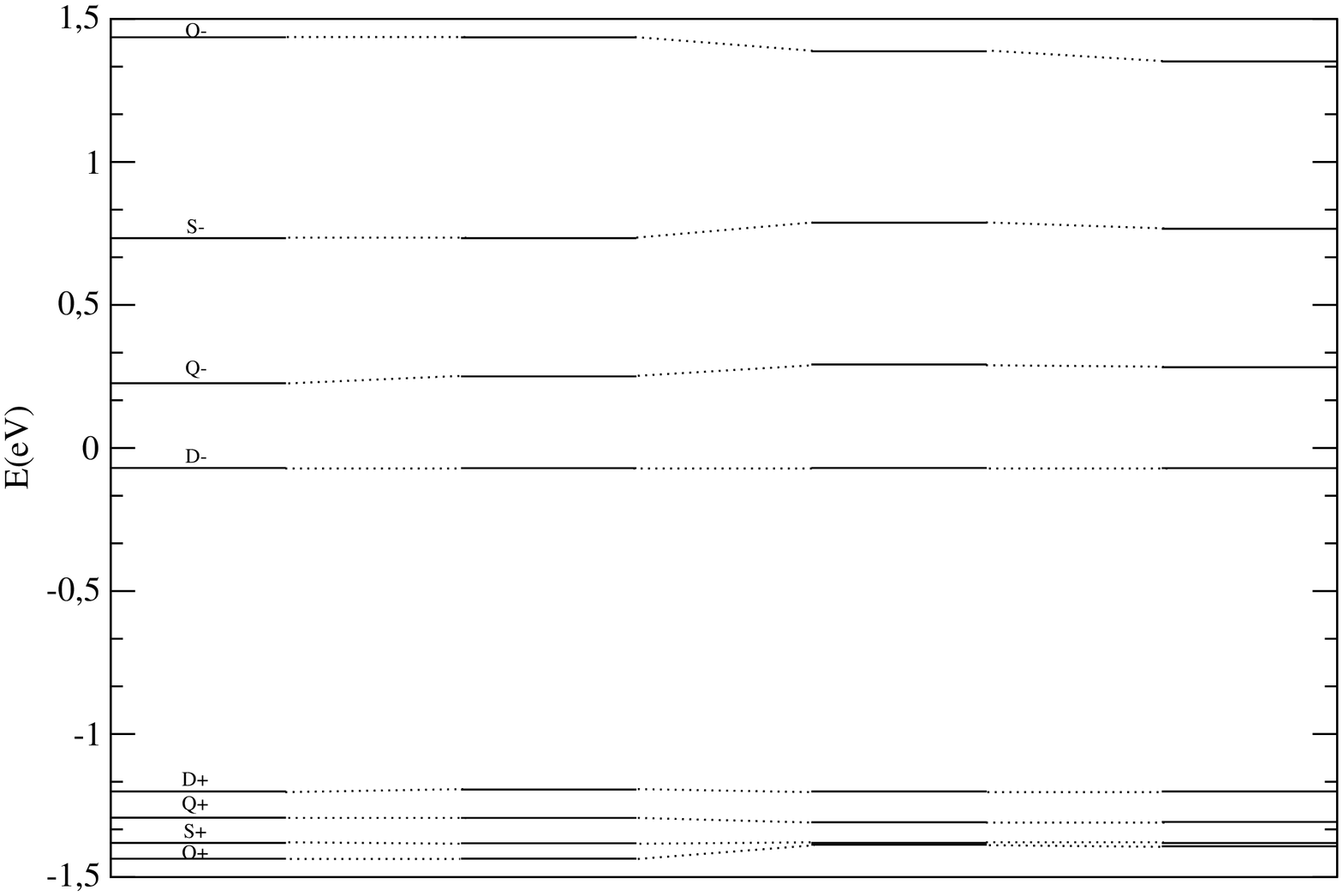}}}\caption{\label{figure2} Comparison of the MS-CASPT2 spectrum of the embedded cluster (column 1) with the outcomes of the refined DE model (column 2), the usual double exchange model (colum 3) and the Heisenberg model (column 4). } \end{figure}

Taking the CASSCF wave functions as references, we determined the second-order perturbation theory MS-CASPT2 spectrum 
for the low-lying states of the cluster. 
The ordering of the states is the following:
\begin{widetext}
\begin{equation}
E(O_-)<E(S_-)<E(Q_-)<E(D_-)<E(D_+)<E(Q_+)<E(S_+)<E(O_+),
\end{equation}
\end{widetext}
which is compatible with all the here-considered models. 
The electronic interactions of the different models optimized in order 
to reproduce at best the \textit{ab initio} spectrum are given in Table \ref{table1} 
and the corresponding model spectra are represented in Figure \ref{figure2} 
together with the MS-CASPT2 spectrum. Using Eq. \ref{extractt}, we validate the numerical fitting procedure. Unfortunately, the comparison cannot be done at the MS-CASPT2 level since this method does not provide the necessary wave function coefficients.\cite{footnote} Instead, we take the spectrum provided by the zeroth-order wave function and extract $t$ both by fitting the usual DE Hamiltonian and by applying Eq. \ref{extractt}. The respective values of 0.9043 eV and 0.9108 eV compare very well.

From the comparison of the different spectra, it appears that 
the qualitative features of the \textit{ab initio} spectrum are reproduced in all model spectra, legitimating 
the consideration of the three different models. Notice that this is not the case for all Hamiltonians: The mechanism proposed by Zener,\cite{Zen} which only considers the hopping integral and does not introduce any antiferromagnetic contribution, cannot reproduce the \textit{ab initio} spectrum.\cite{nous1}

Concerning the extracted electronic interactions 
and the mean error per state of the different models, several comments can be made. 
\begin{itemize}
\item The mean error per state of the usual DE and Heisenberg models 
are similar but not strictly identical as in the symmetric cluster.\cite{nous1,nous2}
\item As already observed for symmetric DE compounds, the explicit consideration of the non-Hund 
atomic states improves quantitatively the reproduction of the \textit{ab initio} 
spectrum and therefore the modelization.
An error of only 0.14\% is obtained for the refined DE model. 
\item The non-symmetry of the cluster (which is responsible for the charge ordering) 
is reflected in the difference of the $J_1$ and $J_3$ exchange integral values 
of the Heisenberg model. The very large value of $J_3$  shows the covalent character of the interaction 
between Mn$_b$ and the bridging O$_c$ oxygen. Actually, the value is so large that it even questions the validity 
of a purely magnetic description of the cluster.
\item The values of the electronic interactions $t$ and $J$
extracted from the two DE models are qualitatively different. 
In particular, the value of the exchange integral $J$ 
of the usual DE model is twice as large as the one extracted from the 
refined DE model. A large difference between the two extracted values
has already been observed in other 
studies.\cite{nini,rere,nous1,nous2} The participation of non-Hund states in the \textit{ab initio} 
wave function enhances the covalent character of the interaction between the Mn sites. 
Their explicit inclusion in the model space results in an antiferromagnetic contribution, i.e. it stabilizes the low and intermediate spin states. 
In order to reproduce this antiferromagnetic contribution, the usual DE model  
overestimates the value of the antiferromagnetic exchange integral $J$ and underestimates the value of $t$, i.e. of the ferromagnetic contribution. 
\end{itemize}

\begin{table}
\caption{\label{table1}Effective parameters of the refined DE (non-Hund), usual DE (Zener, Girerd Papaefthymiou, ZGP), and Heisenberg model extracted from the MS-CASPT2 spectrum. 
The magnetic exchange ($J$) and hopping ($t$) integrals are in eV.
$q_{min}$ is the value of the effective parameter $q$ 
at the minimum of the ground state potential energy curve, $\Delta $ (in eV) is the curvature of the potential energy curves of the diabatic (i.e. before interaction) 
left and right states. 
$\epsilon $ is the mean error of the fit of the model Hamiltonian spectrum to the \textit{ab initio} spectrum.}
\begin{ruledtabular}
\begin{tabular}{ccccc}
                          &  non-Hund    &      ZGP      &   Heisenberg   \\
     $J_1$                &              &                 &   0.568       \\
     $J_3$                &              &                 &   0.939       \\
     $J_2$                &              &                 &   0.082       \\
      $J$                 &   0.0405     &  0.0847         &               \\
     $\delta_H $            &   4.993     &                 &               \\
$\Delta $                 &   0.7177     &  0.6935         &               \\
$q_{min}$                &   $\pm$0.38       & $\pm$0.32           &               \\
       $t$                &   1.3292     &  1.3081         &               \\
$\epsilon $               &   0.14       &  1.014          &    1.010      \\
\end{tabular}
\end{ruledtabular}
\end{table}

\subsection{Role of the Non-Hund states on the spin multiplicity of the ground state}

As already shown in the peculiar case of a symmetric cluster,\cite{nous1,nous2} both the exchange integral $J$ which describes the electronic circulation in the $t_{2g}$-like orbitals and the non-Hund states stabilize the low and intermediate spin states with respect to the highest spin one. If these antiferromagnetic contributions are large enough they may even generate an intermediate spin ground state.\cite{rere,david} We have seen in the previous subsection that the model spectra present the same qualitative features as the \textit{ab initio} spectrum at the experimental geometry.

\begin{figure}[t]
\centerline{\rotatebox{0}{\includegraphics[scale=0.6]{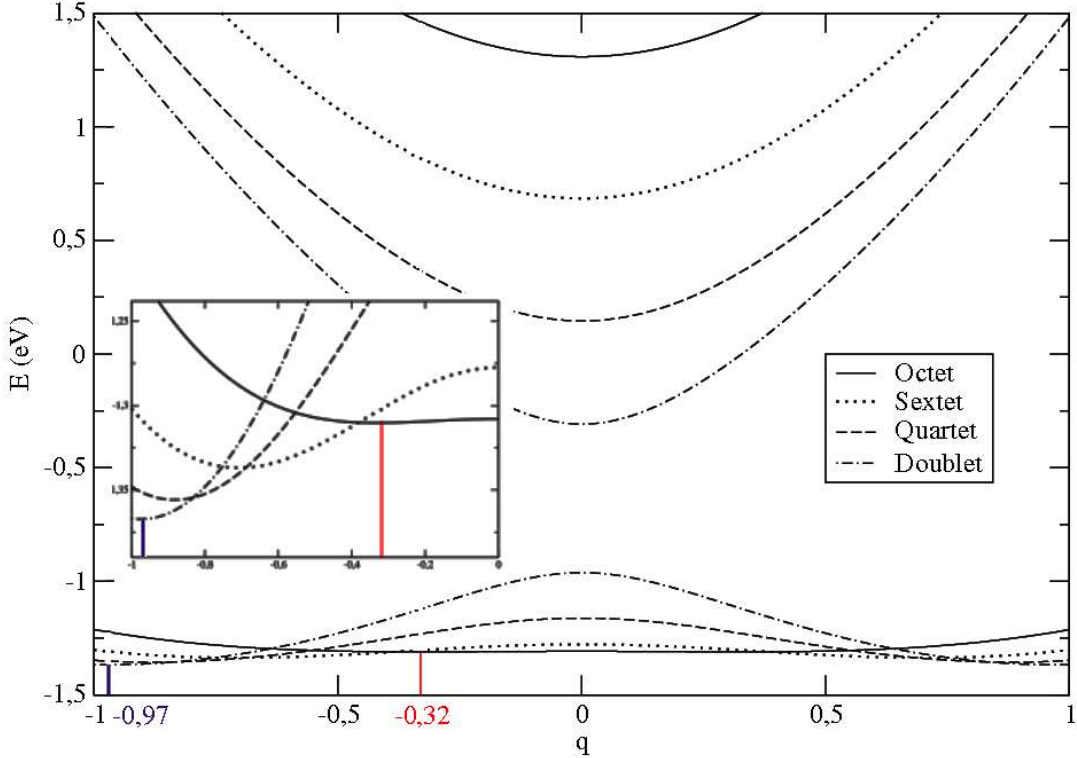}}}\caption{\label{figure3} Potential energy curves of the eight states obtained from the optimized parameters of the usual DE model. The red line indicates the optimized value of q$_{min}$ in the octet ground state i.e. the geometry for which the  \textit{ab initio } spectrum has been calculated. The blue line indicates the position of the doublet state well.}
\end{figure}

\begin{figure}[t]
\centerline{\rotatebox{-90}{\includegraphics[scale=0.6]{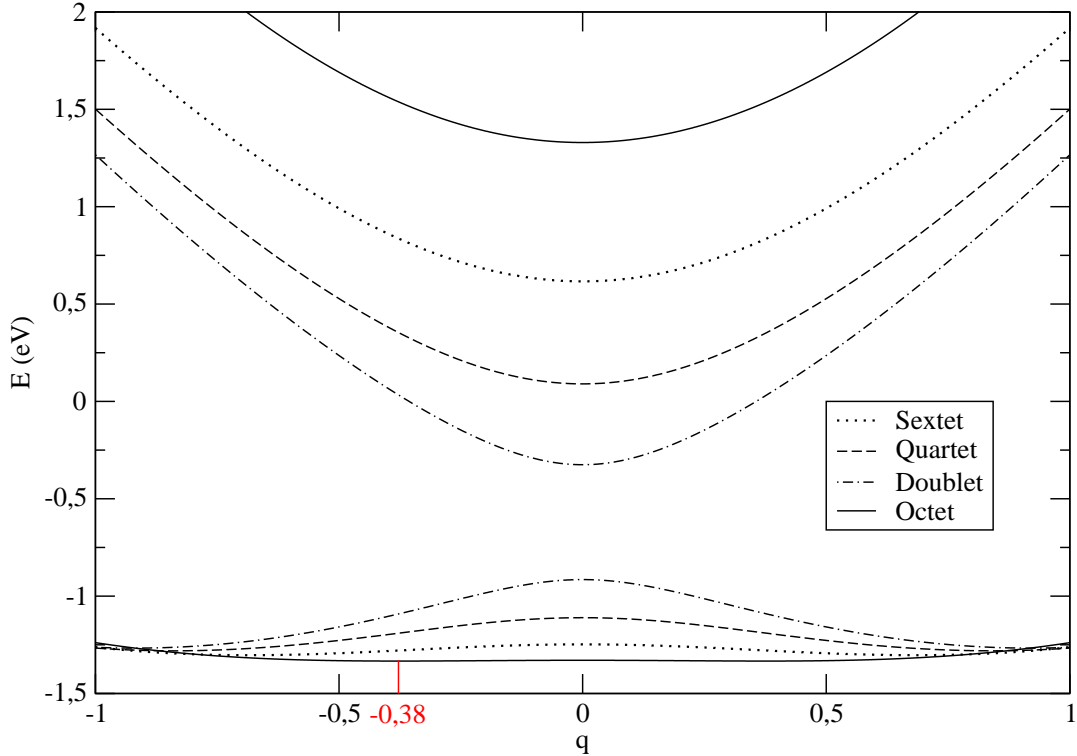}}}\caption{\label{figure4} Potential energy curves of the eight states obtained from the optimized parameters of the refined DE model. The red line indicates the optimized value of q$_{min}$ in the octet ground state}
\end{figure}

As shown in Sec. \ref{model2}, the curvature $\Delta$ of the potential energy curves of the diabatic spin states can be extracted from the calculated spectrum. From the value of $\Delta$ and the optimized effective electronic interactions, it becomes possible to determine the full local potential energy curves of the adiabatic states as functions of the coordinate $q$ for the two DE models. Figures \ref{figure3} and \ref{figure4} depict the curves of the eight spin states for the usual and the refined DE model, respectively. It is striking to see that while the octet state is the ground state at the experimental geometry in the usual DE model, the potential energy curves show more stable minima for low and intermediate spin states at larger values of $q$. On the contrary, the octet potential energy curve is the lowest state over the whole interval for the refined DE model.

This qualitative disagreement between the usual DE model and the \textit{ab initio} calculation is due 
to the overestimation of the exchange integral $J$. In absence of the hopping integral $t$
(i.e. for the diabatic states), the potential wells associated with the left and right localized holes 
are such that the ordering of the eight lowest electronic states is governed by the Heisenberg contribution. This antiferromagnetic interaction induces an antiferromagnetic order and makes the doublet state the ground state. From Eq. 4 and 7 it is easily seen that the coupling between the left and right diabatic states 
through the hopping integral 
is more effective in the highest spin state. Hence,
this interaction, which is the largest interaction of the DE mechanism,
should induce a high spin ground state.
However, the artificially large $J$ makes that the hopping interaction is not large enough 
to restaure a ferromagnetic order in the usual DE model.
As a consequence, the lowest potential energy well is the doublet state one. The minimum of this potential energy curve lies at  $q_{min}$=$\pm$0.97, 0.06 eV below the minimum in the octet curve at $\pm$0.32.
Owing to the antiferromagnetic contribution of the non-Hund states 
in the refined DE model, the optimized exchange integral $J$ is properly 
estimated and both the spectrum at the experimental geometry and 
the potential energy curves of the different states reproduce the 
correct physics of the system.

\section{Conclusions}

The extraction of local electronic structure parameters in the CE-phase of the half-doped manganite La$_{0.5}$Ca$_{0.5}$MnO$_3$ reveals that the local physics can accurately be described with a model Hamiltonian that includes the usual double exchange physics of mixed valence compounds plus an additional term due to the non-Hund states. The large $t$-value (1.3 eV) indicates a rather mobile electron and hence small charge disproportionation. In fact, Mulliken population analysis associates a slightly higher charge to the Mn-ion in the Jahn-Teller distorted site. The difference between both sites is 0.07 electron. Although absolute Mulliken charges are not reliable, the relative values can be considered as indicative for tendencies. Moreover, the observed charge disproportionation of 0.07 is consistent with the maximum of 0.18 predicted in model Hamiltonian studies.\cite{vandenBrink:1999}

The PKS model provides a simple recipe to obtain the potential energy curves of the low-lying states as function of the distortion parameter $q$ and hence obtain more detailed information about the local electronic physics. The curves for the low-lying states show a double well, where the minimum of the high spin ground state curve lies at q $\pm$ 0.38. This corresponds to a situation in between the completely delocalized solution with $q_{min}$~=~0 and the completely localized solution for $q_{min}$=$\pm$1. The latter solution has a disproportionation of 1 electron while for the $q_{min}$=0 situation there is no barrier and the charge disproportionation has completely disappeared.\cite{footnote2}

The height of the barrier between left and right localized solutions is 0.04 eV for the octet ground state of the embedded cluster. Our model study only considers local distortions around the two Mn-sites considered in the cluster and the collective nature of the distortions in the CE-phase could lead to higher barriers. Obviously these effects should be included in order to be conclusive about the nature of the local electronic structure of the half-doped manganite; charge ordered or delocalized (Zener polaron) as proposed in Ref. \onlinecite{daoud}. Nevertheless, the smallness of the barrier is in line with the two different interpretations obtained by experimentalists and the controversy in the literature.

\begin{acknowledgments}
Financial support has been provided by the Spanish Ministry of Education and Culture under Project No. CTQU2005-08459-C02-02/BQU, and the Generalitat de Catalunya (grant 2005SGR-00104).
\end{acknowledgments}


\begin{thebibliography}{9999}
\bibitem{footnote} Due to the so-called contracted nature of the first-order corrected wave function, the relative size of the coefficients of the determinants that belong to the CAS are not corrected for the dynamical electron correlation effects.
\bibitem{footnote2} Applying the PKS model to the Zener polaron structure leads to a $q_{min}$ that is very close to zero, indicating the absence of charge disproportionation for this structure.
\end{thebibliography}
\end{document}